\documentclass[12pt]{article}

\usepackage{times}
\usepackage{geometry} 
\usepackage{times}
\usepackage{theorem}
\usepackage{ifthen}
\usepackage{graphicx}
\usepackage{amsmath}
\usepackage{amssymb}
\usepackage{mathrsfs}
\usepackage{enumerate}

\newcommand\thmheadfont{\upshape\bfseries}
\theoremheaderfont{\thmheadfont}
\theorembodyfont{\itshape}
\newtheorem{thm}{Theorem}
\theoremstyle{plain}
\theorembodyfont{\upshape}

\newtheorem{lemma}[thm]{Lemma}

\newcommand{\hole}{\Box}

\newenvironment{pr}[1][!*!,!]%
{\noindent{\thmheadfont Proof%
    \ifthenelse{\equal{#1}{!*!,!}}{}{%
      \normalfont\ (#1)}\\}}%
{$\square$\medskip}
{\noindent{\thmheadfont Proof%
    \ifthenelse{\equal{#1}{!*!,!}}{}{%
      \normalfont\ (#1)}\ \ \ }}%
{$\square$
\medskip}

\title{Forest Categories}
\author{ Howard Straubing\\Boston College}

\begin{document}

\maketitle

\begin{abstract}  

A growing body of research into the expressive power of logics on trees has employed algebraic methods---especially the {\it syntactic forest algebra}, a generalization of the syntactic monoid of regular languages.  Here we enlarge the mathematical foundations of this study by extending Tilson's theory of the algebra of finite categories, and in particular, the Derived Category Theorem, to the setting of forest algebras.  As an illustration of the usefulness of these methods, we provide a new treatment of the recent results of Place and Segoufin on locally testable tree languages.
\end{abstract}
\section{Introduction}

\begin{quote}
Note added, January 2018:  This paper was written in 2011, submitted to a conference, and rejected.  I made no further effort to publish it, since although the extension of Tilson's theory to the forest algebra setting is formally correct, I was never persuaded of the usefulness of the method, and not happy with the rather complicated application given here.  Recently, Michael Hahn and Andreas Krebs dusted off this old work of mine and applied to a quite different problem, so I thought it was a good time to make it publicly available.  I have made no improvements or modifications---aside from this note---in the past seven years. Like an old house that needs a lot of work, I offer the paper `as is'.

\end{quote}
 While it might not appear so at first, this paper is part of an ongoing research effort to understand the expressive power of various predicate and temporal logics on trees.  Analogous problems for words, rather than trees, have been studied for well over forty years, and much of this research has relied on algebraic methods.  Typically, the set $L$ of words that satisfies a formula in one of the logics under consideration is a regular language, and expressibility in the logic is reflected in properties of the syntactic monoid or the syntactic morphism of $L.$  A large compendium of results in this vein (up to 1994) is presented in Straubing ~\cite{straubing} for various predicate logics.  Additional results, mostly oriented around temporal logic are in Th\'erien and Wilke~\cite{tw} and Wilke~\cite{wilke}.
 
 Recently this approach has been extended, with some success, to languages of trees, especially unranked trees.  Here, the {\it syntactic forest algebra,} introduced by Bojanczyk and Walukiewicz~\cite{bw} plays the role of the syntactic monoid.  A number of papers have appeared that use this construct either explicitly (Bojanczyk, {\it et.al.}~\cite {bsw,bss,bs,b}) or implicitly (Benedikt and Segoufin~\cite{ben-seg}, Place and Segoufin~\cite{ps}) to characterize the languages definable in certain tree logics.  A related approach for trees of bounded rank is described in Esik and Weil~\cite{ew1,ew2}.
 
 It was recognized early on in the development of the theory for words that wreath product decompositions play an important role .  Thus McNaughton and Papert~\cite{mp} turned to the Krohn-Rhodes decomposition for the characterization of properties definable in first-order logic, and Brzozowski and Simon~\cite{brz-sim} studied decompositions with an idempotent and commutative factor in their work on locally testable languages.  A close reading of this latter work, along with related papers, led to a more general understanding of the role played by graph congruences in obtaining these decompositions, initially in Straubing~\cite{v*d} and Th\'erien and Weiss~\cite{tw}.  This reached its definitive form in Tilson's work~\cite{tilson} on the algebra of finite categories.  An up-to-date account appears in the book by Rhodes and Steinberg~\cite{rs}.  Categories have emerged as an important tool in the study of finite semigroups and the languages they accept, one that has found applications well beyond the problems that originally inspired it.  (See, {\it e.g.,} Pin, {\it et. al.} ~\cite{pst,pin}.)
 
 It is now known that wreath products of forest algebras also figure importantly in the study of logics on trees.  This is seen in the many examples studied in Bojanczyk, {\it et. al.}~\cite{bsw}, which include first-order logic with the ancestor relation, and the temporal logics $CTL$ and $CTL^*.$ It also underlies much of what is done in ~\cite{ps}. So the time is ripe for developing new algebraic tools for producing such decompositions.
 
 In the present paper we extend the algebra of finite categories from monoids to forest algebras.  After reviewing the basics of forest algebras  in Section 2, we give the definition of forest categories and of the derived category of a pair of morphisms.  The tricky part here is in getting the definitions just right.  Once that is done, it is a simple matter to establish an analogue to the Derived Category Theorem of Tilson~\cite{tilson}, which connects categories to wreath products.  In the sections that follow, we endeavor to show that this is more than just abstract nonsense.  In Section 5 we establish efficient necessary and sufficient conditions for determining if a given forest category divides an idempotent and commutative forest algebra---an analogue to the result of Simon~\cite{brz-sim} that started it all.  We then show how this underlies the very recent work on locally testable forest languages by Place and Segoufin~\cite{ps}.  It is this paper that provided much of the inspiration for the present work.  However we cannot stress enough that what we propose here is more than just a new proof of the results in ~\cite{ps}, or, worse yet, the same proof couched in an obscure new language.  Rather, we are introducing a new mathematical tool, which like its precursor for monoids, promises to have many more applications than the one we have chosen as a proof of concept.  We discuss the prospects for these in the last section.

For the usual reasons of space, the three most detailed and technical proofs are omitted.  However it should be noted that all three of these (the proofs of Theorems~\ref{theorem.derived_category} and \ref{theorem.local_global_ic}, and Lemma~\ref{lemma.delay_theorem}) closely follow arguments that already appear in the literature.

 
\section{Forest Algebras}

For more background on forest algebras, see Bojanczyk, {\it et. al.} ~\cite{bw, bsw}. A forest algebra is little more than a pair of monoids, where one of the monoids (the {\it vertical} monoid) acts on the underlying set of the other (the {\it horizontal} monoid).  In this paper we depart from the tradition that has begun to be established in the articles that have appeared on forest algebras---and return to the much longer-established traditions of semigroup theory---by writing the action as a right action.  We do this precisely because we are generalizing the algebra of finite categories, and we would be forced to reverse the normal left-to-right direction of arrows in categories in order to accommodate left actions in forest algebras.

Here is the precise definition:  A {\it forest algebra} is a pair $(H,V)$ of monoids with some additional properties, which we will specify shortly. The operation in $H$ is written additively and the identity written 0. In this paper we will also suppose that $H$ is commutative.  This is not usually a requirement in discussions of forest algebras, and our theory of forest categories will probably work just fine without it, but it does simplify the presentation and will serve for our purposes. The operation in $V$ is written multiplicatively, and its identity is denoted 1.  $V$ acts on $H$ on the right, so that given $v\in V,$ $h\in H$ there is an element $hv$ of $H.$  The properties that make this an action are {\it (i)} $(hv_1)v_2=h(v_1v_2)$ if $h\in H$ and $v_1,v_2\in V,$ and {\it (ii)} $h1=h$ for all $h\in H.$  We require this action to be {\it faithful,} which means that if $hv=hv'$ for all $h\in H,$ then $v=v'.$  There is just one additional property in the definition:  If $h\in H$ and $v\in V,$ then there is an element $ins(v,h)$ of $V$ such that for all $g\in H,$ $g\cdot ins(v,h)=gv+h.$  We usually write the more natural-looking $v+h$ instead of $ins(v,h).$  Observe that the map $h\mapsto 1+h$ embeds the monoid $H$ into $V.$

A homomorphism $\alpha:(H,V)\to (H',V')$ of forest algebras is actually a pair of monoid homomorphisms $\alpha_H:H\to H',$ $\alpha_V: V\to V',$ with the additional property that $\alpha_H(h)\alpha_V(v)=\alpha_H(hv)$ for all  $h\in H,$ $v\in V.$ We usually drop the subscript and write $\alpha$ for both components.  Notice that a homomorphism in this sense automatically preserves the $ins$ operation, and thanks to our various notational conventions we very conveniently have $\alpha(h+v)=\alpha(h)+\alpha(v)$ for any $h\in H,$ $v\in V.$ We say a forest algebra $(H,V)$ {\it divides} a forest algebra $(H',V'),$ and write $(H,V)\prec (H',V')$ if $(H,V)$ is a homomorphic image of a subalgebra of $(H',V').$  
Given two forest algebras $(H_1,V_1),$ $(H_2,V_2)$ we define the {\it wreath product}
$$(H_2,V_2)\circ (H_1,V_1)=(H_2\times H_1,V_2^{H_1}\times V_1)$$
exactly as one defines the wreath product of transformation monoids: The action is given by
$$(h_2,h_1)(f,v_1)=(h_2\cdot f(h_1), h_1v).$$
The monoid structure on $H_2\times H_1$ is simply the direct product.  The multiplication in the vertical monoid is defined by
$$(f_1,v_1)(f_2,v_2)=(g,v_1v_2),$$
where for all $h\in H_1,$ $g(h)=f_1(h)f_2(hv_1).$  It is straightforward to show  the wreath product is a forest algebra.  The pair of maps projecting onto the right-hand coordinates is a forest algebra homomorphism $\pi.$

Let $A$ be a finite alphabet.  We describe the {\it free forest algebra}Ê $A^{\Delta}=(H_A,V_A)$ as follows:  $H_A$ consists of expressions built starting with 0 and closing under adjunction of a letter $a\in A$ on the right and under $+.$  An example of such an expression, with $A=\{a,b\},$ is
$$((0\cdot a)b+ (0\cdot a+0\cdot b))a+(0\cdot b)a .$$
We will usually drop the 0's that appear in such an expression, as well as the parentheses whenever the action axioms make these redundant, so we will write this more simply as
$$(ab+a+b)a+ba,$$
and depict it in the obvious fashion as a forest with two trees, both with $a$ at the root and with four leaves altogether, two labeled $a$ and two labeled $b.$  Since we are assuming $H_A$ is commutative, we identify many different forests:  For instance, the one above is identical to 
$$ba+(a+ab+b)a.$$
$V_A$ consists of {\it contexts}:  These are forests in which one of the leaves has been removed and replaced by a hole:  For example
$$(ab+a+b)a+\hole a.$$
Contexts act on forests by substituting a forest $s\in H_A$ for the hole in a context $p\in V_A$ to form a forest $sp.$  Contexts are composed by substituting $p\in V_A$ for the hole in $q\in V_A$ to form a context $pq.$  As a result $A^{\Delta}$ is indeed a forest algebra.  What makes it the `free' forest algebra is this universal property:  If $(H,V)$ is a forest algebra, then any map $f:A\to V$ extends to a unique forest algebra homomorphism $\alpha:A^{\Delta}\to (H,V)$ such that for all $a\in A,$ $\alpha(\hole a)=f(a).$

A subset $L$ of $H_A$ is called a {\it forest language.}  A forest language $L$ is {\it recognized} by a forest algebra $(H,V)$ if there is a homomorphism $\alpha:A^{\Delta}\to (H,V)$ such that $L=\alpha^{-1}(X)$ for some $X\subseteq H.$  A forest language is {\it regular} if it is recognized by a finite forest algebra---this is equivalent to recognition by a bottom-up deterministic automaton.   For every regular forest language $L,$ there is a unique minimal algebra $(H_L,V_L)$ recognizing $L,$ in the sense that for every forest algebra $(H,V)$ recognizing $L,$ $(H_L,V_L)\prec (H,V).$  $(H_L,V_L)$ is called the {\it syntactic forest algebra} of $L,$ and the homomorphism $\alpha_L:A^{\Delta}\to (H_L,V_L)$ that recognizes $L$ is called the {\it syntactic morphism} of $L.$  Both the syntactic monoid and syntactic morphism are effectively computable from any automaton that recognizes $L.$

   \section{Forest Categories and Division}

A {\it forest category} ${\cal C}$ is a triple $({\bf Obj}({\cal C}),{\bf Arr}({\cal C}),{\bf HArr}({\cal C}))$ of sets with the following properties:
\begin{enumerate}[(a)]
\item  ${\bf Obj}({\cal C})$ is a finite set, called the set of {\it objects} of ${\cal C}.$
\item For all $(x,y)\in {\bf Obj}({\cal C})\times {\bf Obj}({\cal C}),$ there is a set ${\bf Arr}(x,y),$ called the set of {\it arrows} from $x$ to $y,$ such that ${\bf Arr}({\cal C})$ is the disjoint union
$${\bf Arr}({\cal C})=\bigcup_{x,y\in{\bf Obj}({\cal C})}{\bf Arr}(x,y).$$

We denote an arrow $u$ from $x$ to $y$ as $x\stackrel{c}{\rightarrow} y.$  We write $start(u)$ for the object $x,$ and $end(u)$ for the object $y.$

\item For all $x\in {\bf Obj}({\cal C}),$ there is a set ${\bf HArr}(x),$ called the set of {\it  half-arrows} to $x,$ such that ${\bf HArr}({\cal C})$ is the disjoint union
$${\bf HArr}({\cal C})=\bigcup_{x\in{\bf Obj}({\cal C})}{\bf HArr}(x).$$

We denote a  half-arrow $u$  to $x$  as $ \stackrel{c}{\rightarrow} x,$ and we write $end(u)$ for $x.$  

\item ${\bf Obj}(\cal{C})$ is a commutative monoid whose operation is written $+$ and whose identity is written 0. As was the case with our treatment of forest algebras, commutativity is not really a requirement, but it makes the presentation somewhat simpler, and all the applications we discuss here will result in forest categories that are commutative in this sense. 

\item ${\bf HArr}({\cal C})$ is also a commutative monoid, with operation similarly denoted $+,$ and ${\bf HArr}(x)+{\bf HArr}(y)\subseteq{\bf HArr}(x+y),$ so that the map
$ c\mapsto end(c)$
is a monoid homomorphism from ${\bf HArr}({\cal C})$ onto ${\bf Obj}({\cal C}).$ We will often depict the half-arrow  $ \stackrel{c}{\rightarrow} x+ \stackrel{d}{\rightarrow} y$
as

\[
\begin{array}{cc}
 \stackrel{c}{\rightarrow} & x\\
 \stackrel{d}{\rightarrow} & y.\,\\

\end{array}
\]
but will always bear in mind that this is equivalent to a half-arrow $\stackrel{e}{\rightarrow} x+y.$

\item For all $x,y,z\in{\bf Obj}({\cal C})$ there is a binary operation
$${\bf Arr}(x,y)\times{\bf Arr}(y,z)\to{\bf Arr}(x,z).$$
We denote this operation multiplicatively:
$$x\stackrel{c}{\rightarrow} y\cdot y\stackrel{d}{\rightarrow} z\quad=\quad x\stackrel{e}{\rightarrow} z,$$
or sometimes simply as
$$x\stackrel{c}{\rightarrow} y\stackrel{d}{\rightarrow} z=x\stackrel{e}{\rightarrow} z.$$
This operation is associative in the following sense:  for all
$$w\stackrel{c}{\rightarrow} x, x\stackrel{d}{\rightarrow} y, y\stackrel{e}{\rightarrow} z\in {\bf Arr}({\cal C}),$$
$$(w\stackrel{c}{\rightarrow} x \cdot  x\stackrel{d}{\rightarrow} y) \cdot y\stackrel{e}{\rightarrow} z=
w\stackrel{c}{\rightarrow} x \cdot  (x\stackrel{d}{\rightarrow} y \cdot y\stackrel{e}{\rightarrow} z).$$
Further, for each $x\in{\bf Obj}({\cal C}),$ there exists $x\stackrel{1_X}{\rightarrow} x \in{\bf Arr}(x,x)$ such that for all 
$w\stackrel{c}{\rightarrow} x, x\stackrel{e}{\rightarrow} z\in {\bf Arr}({\cal C}),$
$$w\stackrel{c}{\rightarrow} x\cdot x\stackrel{1_X}{\rightarrow} x=w\stackrel{c}{\rightarrow} x,$$
$$x\stackrel{1_X}{\rightarrow} x\cdot  x\stackrel{e}{\rightarrow} z=x\stackrel{e}{\rightarrow} z.$$

\item For all $x,y\in{\bf Obj}({\cal C})$ there is a binary operation
$${\bf HArr}(x)\times{\bf Arr}(x,y)\to{\bf HArr}(y).$$
We denote this operation multiplicatively:
$$\stackrel{c}{\rightarrow} x\cdot\ x\stackrel{d}{\rightarrow} y=\stackrel{e}{\rightarrow} y,$$
or sometimes simply as
$$\stackrel{c}{\rightarrow} x\stackrel{d}{\rightarrow} y=\stackrel{e}{\rightarrow} y.$$
This operation is an action in the following sense:  for all
$\stackrel{c}{\rightarrow} x\in{\bf HArr}({\cal C}),$ $x\stackrel{d}{\rightarrow} y,$ $y\stackrel{e}{\rightarrow} z\in {\bf Arr}({\cal C}),$ 
$$(\stackrel{c}{\rightarrow} x \cdot  x\stackrel{d}{\rightarrow} y) \cdot y\stackrel{e}{\rightarrow} z=
\stackrel{c}{\rightarrow} x \cdot  (x\stackrel{d}{\rightarrow} y \cdot y\stackrel{e}{\rightarrow} z).$$
Further,  for all 
$\stackrel{c}{\rightarrow} x\in {\bf HArr}({\cal C}),$
$$\stackrel{c}{\rightarrow} x\cdot x\stackrel{1_X}{\rightarrow} x=\stackrel{c}{\rightarrow} x.$$
We require this action to be {\it faithful} in the sense that if 
$$\stackrel{c}{\rightarrow}h_2\stackrel{d}{\rightarrow}h_2'=\stackrel{c}{\rightarrow}h_2\stackrel{d'}{\rightarrow}h_2'$$
for all $\stackrel{c}{\rightarrow}h_2\in{\bf HArr}(h_2),$ then
$$h_2\stackrel{d}{\rightarrow}h_2'=h_2\stackrel{d}{\rightarrow}h_2'.$$

\item For each $x\stackrel{c}{\rightarrow} y\in{\bf Arr}(x,y),$ and each $\stackrel{d}{\rightarrow} z\in{\bf HArr}({\cal C})$ there exists
$$ins(x\stackrel{c}{\rightarrow} y,\stackrel{d}{\rightarrow} z)\in{\bf Arr}(x,y+z)$$
such that for all $w\stackrel{f}{\rightarrow} x\in {\bf Arr}({\cal C}),$
$$w\stackrel{f}{\rightarrow} x\cdot ins(x\stackrel{c}{\rightarrow} y,\stackrel{d}{\rightarrow} z)=ins(w\stackrel{f}{\rightarrow} x\cdot x\stackrel{c}{\rightarrow} y,\stackrel{d}{\rightarrow} z),$$
and for all $\stackrel{g}{\rightarrow} v\in{\bf HArr}({\cal C}),$

$$\begin{array}{ll}ins(ins(x\stackrel{c}{\rightarrow} y,\stackrel{d}{\rightarrow} z),\stackrel{g}{\rightarrow} v)&=\\
ins(x\stackrel{c}{\rightarrow} y,\stackrel{d}{\rightarrow} z+\stackrel{g}{\rightarrow} v ).&\\
\end{array}
$$
These axioms are more readable, and look more natural, if we write $ins(x\stackrel{c}{\rightarrow} y,\stackrel{d}{\rightarrow} z)$
as $x\stackrel{c}{\rightarrow} y+\stackrel{d}{\rightarrow} z,$ and depict it as
\[
\begin{array}{ccc}
 x&\stackrel{c}{\rightarrow} & x\\
 &\stackrel{d}{\rightarrow} & y.\,\\

\end{array}
\]

\end{enumerate}

\noindent{\bf Remarks on the Definition}

\begin{enumerate}[(a)]
\item If we leave out everything having to do with half-arrows, this is simply the standard definition of a category with a finite set of objects.

\item A category with a single object is a monoid.  Similarly, in a forest category with only one object, the operations described in parts {\it (f),(g),(h)} of the definition are always defined, and their properties reduce to the axioms for a forest algebra.  Thus a forest category with one object is a forest algebra.

\item Suppose we have a multiset of arrows and half-arrows in a forest category ${\cal C}.$  We can compose these in any fashion that makes the endpoints match up correctly, and obtain a  {\it forest diagram}, as illustrated in Figure 1.  Such a diagram is simply a forest in which the leaf nodes are labeled by half-arrows and the internal nodes by arrows, with the constraint that the start object of each internal node must equal the sum of the end objects of its children.

\begin{figure}[htb]
      \begin{center}
	\includegraphics[width=3.0in,height=1.5in,clip=true]{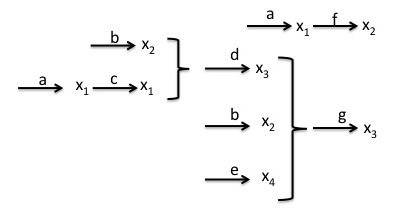}
	\caption{A Forest Diagram}
     \end{center}
    \end{figure} 
    
    In the figure, the arrow labeled $d$ belongs to ${\bf Arr}(x_1+x_2,x_3),$ and the arrow labeled $g$ to ${\bf Arr}(x_2+x_3+x_4,x_3).$  We can view the diagram as a graphical representation of the expression
    $$\begin{array}{l}\stackrel{a}{\rightarrow}x_1\stackrel{f}{\rightarrow}x_2+\\
    \stackrel{a}{\rightarrow} x_1(x_1\stackrel{c}{\rightarrow}x_1+\stackrel{b}{\rightarrow} x_2)\cdot\\(x_1+x_2\stackrel{d}{\rightarrow} x_3 +
    \stackrel{b}{\rightarrow} x_2+\stackrel{e}{\rightarrow} x_4)\cdot x_2 \stackrel{g}{\rightarrow} x_3.\\
    \end{array}$$ 
    
There are other ways to parse the diagram, but thanks to the forest category axioms, all the resulting expressions have the same value in ${\bf HArr}(x_2+x_3).$  That is, any forest diagram $D$  over ${\cal C}$ unambiguously determines an element  $val(D)$ of ${\bf HArr}(x),$ where $x=rootsum(D)$ is the sum of the rightmost objects in the diagram. If the underlying forest of such a diagram has just one component, we will call it a {\it tree diagram}.  

Similarly, if we eliminate one of the half-arrows from the diagram, leaving a single object $y$ exposed at a leaf, we obtain a diagram $D'$ that unambiguously determines an element $val(D')$ of ${\bf Arr}(y,rootsum(D')).$  We call such a diagram a {\it context diagram,} and denote the exposed object $y$ by $start(D').$  In this setting, the action of an arrow on a half-arrow corresponds to the action of a forest on a context.  We just have to make sure that the endpoints match up:  that is, we can plug a forest diagram $D$ into a context diagram $D',$ and obtain a forest diagram $DD',$ so long as $rootsum(D)=start(D').$  In a like manner, the composition of two arrows corresponds to plugging one context diagram into another. \end{enumerate}

\noindent{\bf Division}

In the theory for monoids, the notion of a category ${\cal C}$ {\it dividing} a monoid $M$ plays a crucial role.  The idea is this:  Suppose we want to evaluate the composition of a sequence of arrows
$$x_0\stackrel{a_1}{\rightarrow}x_1\stackrel{a_2}{\rightarrow}x_2\stackrel{a_3}\rightarrow\cdots\stackrel{a_n}\rightarrow x_n$$
in a category ${\cal C}.$  
We associate to each arrow $x_{i-1}\stackrel{a_i}{\rightarrow}x_i$ 
an element $m_i$ of $M$ in such a manner that knowledge of the terminal objects $x_0$ and 
$x_n$ and of the product $m_1\cdots m_n$ in $M$ is enough to determine the value in ${\bf Arr}({\cal C}).$ There is no problem in this scheme if several different elements of $M$ are associated to the same arrow, or the same element of $M$ to different arrows, so long as no element of $M$ is associated to two distinct arrows with the same endpoints ({\it coterminal} arrows).

For forest categories, the idea is much the same: We want to associate to each half-arrow and arrow of ${\cal C}$ horizontal and vertical elements, respectively, of a forest algebra $(H,V).$  If we associate such a `covering element' to every half-arrow and arrow of a forest diagram $D$ and evaluate the corresponding forest in $(H,V),$ then this value, together with $rootsum(D),$ is enough to determine $val(D),$ and similarly for context diagrams.

Here is the formal definition: If ${\cal C}$ is a forest category and $(H,V)$ a forest algebra, then we write ${\cal C}\prec (H,V),$ and say ${\cal C}$  {\it divides} $(H,V),$ if for each $\stackrel{c}{\rightarrow} x\in{\bf HArr}({\cal C})$ there exists a nonempty set $K_{ \stackrel{c}{\rightarrow} x}\subseteq H,$ and for each $x\stackrel{d}{\rightarrow} y\in{\bf Arr}({\cal C})$ there exists a nonempty set $K_{x\stackrel{d}{\rightarrow} y}\subseteq{\bf Arr}({\cal C})$ satisfying the following properties: 
\begin{enumerate}[(a)]
\item (Preservation of Operations)  For all $\stackrel{c}{\rightarrow} x,$ $\stackrel{d}{\rightarrow} y\in{\bf HArr}({\cal C}),$ $x\stackrel{e}{\rightarrow} y, y\stackrel{f}{\rightarrow} z\in{\bf Arr}({\cal C}) ,$

\begin{enumerate}[(i)]
\item $K_{x\stackrel{e}{\rightarrow} y}\cdot K_{y\stackrel{f}{\rightarrow} z}\subseteq K_{x\stackrel{e}{\rightarrow} y\stackrel{f}{\rightarrow} z}$
\item $K_{\stackrel{c}{\rightarrow} x}\cdot K_{x\stackrel{e}{\rightarrow} y}\subseteq K_{\stackrel{c}{\rightarrow}x\stackrel{e}{\rightarrow} y}$
\item $K_{\stackrel{c}{\rightarrow} x}+ K_{\stackrel{d}{\rightarrow} y}\subseteq K_{\stackrel{c}{\rightarrow} x+\stackrel{d}{\rightarrow} y}$
\item $K_{\stackrel{c}{\rightarrow} x}+ K_{y\stackrel{f}{\rightarrow} z}\subseteq K_{\stackrel{c}{\rightarrow} x+y\stackrel{f}{\rightarrow} z}.$

\end{enumerate}

\item (Injectivity)

\begin{enumerate}[(i)]
\item If $x\stackrel{c}\rightarrow y$ and $x\stackrel{c'}\rightarrow y$ are distinct arrows, then
$K_{x\stackrel{c}\rightarrow y}\cap K_{x\stackrel{c'}\rightarrow y} = \emptyset.$
\item If $ \stackrel{c}\rightarrow y$ and $\stackrel{c'}\rightarrow y$ are distinct half-arrows, then
$K_{\stackrel{c}\rightarrow y}\cap K_{\stackrel{c'}\rightarrow y} = \emptyset.$

\end{enumerate}

\end{enumerate}

If $u$ is either an arrow or a half-arrow, then we say $K(u)$ {\it covers} $u.$
\section{The Derived Forest Category}

Let $A$ be a finite alphabet, and consider a pair of forest algebra homomorphisms

$$(H_1,V_1)\stackrel{\alpha}{\leftarrow}A^{\Delta}\stackrel{\beta}{\rightarrow} (H_2,V_2)$$
mapping {\it onto} finite forest algebras $(H_1,V_1), (H_2,V_2).$  It is a common practice, when dealing with monoids, to view this pair as defining a {\it relational morphism} $\phi=\beta\alpha^{-1}:(H_1,V_1)\to H_2,V_2),$ and work directly with the relation $\phi$ (note that $\phi$ is in general multi-valued and therefore not a homomorphism in the usual sense), however we find it simpler to refer directly to the maps $\alpha$ and $\beta.$  

We define a category ${\cal D}_{\alpha,\beta}$ as follows:

\begin{enumerate}[(a)]
\item ${\bf Obj}({\cal D}_{\alpha,\beta})=H_2.$
\item We set, for $h\in H_2,$
$${\bf HArr}(h)=\{(\alpha(s),h):s\in H_A,\beta(s)=h\}.$$
In other words, ${\bf HArr}(h)$ is the {\it graph} of the relation $\phi.$  We will depict the half-arrow $(h_1,h_2)$ as $\stackrel{h_1}{\rightarrow} h_2.$

\item To define ${\bf Arr}({\cal D}_{\alpha,\beta}),$ we first introduce an equivalence relation on the set
$$\{(h,p,h'): h,h'\in H_2; p\in V_A; h\cdot \beta(p)=h'\}.$$
We define $(h,p,h')\sim (h,q,h')$ if for all $s\in H_A$ with $\beta(s)=h,$ we have $\alpha(sp)=\alpha(sq).$  We then set ${\bf Arr}(h,h')$ to be the set of equivalence classes of $\sim.$  We will still depict an arrow as 
$$h\stackrel{p}{\rightarrow} h',$$
where $p\in V_A,$ but with the understanding that the same arrow has many distinct representations in this form.
\item Note that ${\bf Obj}({\cal D}_{\alpha,\beta})$ and ${\bf HArr}({\cal D}_{\alpha,\beta})$ are commutative monoids, and that the projection of a half-arrow onto its end object is a homomorphism, as required in the definition.  We must now define the other operations in the category and show that they have the desired properties.  We set
$$h_1\stackrel{p}{\rightarrow}h_2\stackrel{q}\rightarrow h_3=h_1\stackrel{pq}{\rightarrow} h_3.$$
Observe that $h_1\beta(pq)=h_1\beta(p)\beta(q)=h_2\beta(q)=h_3,$ so the right-hand side of the above equation is indeed the representation of an arrow.  We still need to show that this is well-defined; in other words, that
$$(h_1,p,h_2)\sim (h_1,p',h_2),$$

$$ (h_2,q,h_3)\sim (h_2,pq,h_3)$$
implies
$$(h_1,pq,h_3)\sim(h_1,p'q',h_3).$$
To this end, let $s\in H_A$ and $\beta(s)=h_1.$  Then the two equivalences imply
$$\alpha(sp)=\alpha(sp'),$$
and, since $\beta(sp)=h_2,$
$$\alpha(spq)=\alpha(spq'),$$
so the two together give
$$\alpha(spq)=\alpha(sp'q'),$$
as required.  Associativity follows at once from associativity in $V_A,$ and the arrow $h\stackrel{1_{V_A}}{\rightarrow} h$ is the identity at $h\in H_2.$

\item  We define the action of an arrow on a half-arrow by

$$\stackrel{h_1}{\rightarrow}h_2\stackrel{p}{\rightarrow}h_2'=\stackrel{h_1\alpha(p)}{\longrightarrow}h_2'.$$
Note that the right-hand side is indeed a half-arrow, since if $\alpha(s)=h_1$ and $\beta(s)=h_2,$ then $\alpha(sp)=h_1\alpha(p)$ and $\beta(sp)=h_2'.$  Furthermore, this operation is well-defined, since if $(h_2,p,h_2')\sim (h_2,q,h_2'),$ then $h_1\alpha(p)=h_1\alpha(q)$ by definition.  The associativity of the action follows directly from the associative law for the action in $(H_1,V_1).$  The definition of equivalent arrows also ensures that this action is faithful.

\item We can set 
$$h\stackrel{p}{\rightarrow} h'+\stackrel{h_1}{\rightarrow}h_2=h\stackrel{p+s} {\rightarrow}h'+h_2,$$
where $s\in H_A$ is such that $\alpha(s)=h_1,$ $\beta(s)=h_2.$  We have 
$$h\beta(p+s)=h\beta(p)+\beta(s)=h'+h_2,$$
 so the right-hand side of the definition represents an arrow, and it is trivial to verify that this is well-defined.  The required algebraic properties follow directly from those for the insertion operation in $(H_1,V_1).$

\end{enumerate}

Our main result connects the derived category to the wreath product.

\begin{thm} (Derived Category Theorem)\label{theorem.derived_category}

Let $A,$ $\alpha,$ $\beta,$ $(H_1, V_1),$ $H_2,V_2)$ be as above, and let $(H,V)$ be a finite forest algebra.
\begin{enumerate}[(a)]
\item If ${\cal D}_{\alpha.\beta}\prec (H,V),$ then
$$(H_1,V_1)\prec (H,V)\circ (H_2,V_2).$$

\item Suppose $\alpha$ factors as 
$$\alpha=\gamma\delta:A^{\Delta}\to (H_1,V_1),$$
where
$$\delta:A^{\Delta}\to (H,V)\circ (H_2,V_2),\gamma:{\rm Im} \delta \to (H_1,V_1),$$
and that $\beta=\pi\delta,$ where $\pi$ is the projection homomorphism from the wreath product onto its right-hand factor.  Then ${\cal D}_{\alpha,\beta}\prec (H,V).$
\end{enumerate}

\end{thm}

The proof is largely a straightforward verification, but there are lot of things to verify, so we give the complete argument in the appendix.

\section{Globally Idempotent and Commutative Forest Categories}

Much of the work in applying categories to automata over words entails finding effective conditions for determining when a finite category ${\cal C}$ divides a monoid belonging to some specified variety of finite monoids.  The earliest such result (which, of course, predates Tilson's introduction of category division, and provided much of the inspiration for the development of the subject), implicit in the work of Brzozowski and Simon~\cite{brz-sim} and McNaughton~\cite{mcn} on locally testable languages, establishes necessary and sufficient conditions for a finite category to divide a finite idempotent and commutative monoid.

\begin{thm}\label{theorem.simon}
A finite category ${\cal C}$ divides a finite idempotent and commutative monoid if and only if for every $x\in {\bf Obj}({\cal C}),$ the monoid ${\bf Arr}(x,x)$ is idempotent and commutative.
\end{thm}

Here we will study an analogous question for forest categories.  Let $H$ be an idempotent and commutative monoid, with its operation written additively.  $H,$ of course, acts faithfully on itself, and the result is a forest algebra $(H,H).$  We call such a forest algebra a {\it flat idempotent and commutative forest algebra,} because when we evaluate the homomorphic image of a forest $s\in H_A$ in $(H,H),$ the value does not depend on the tree structure at all, but only on the node labels.  We say that a forest category ${\cal C}$ is {\it globally idempotent and commutative} if it divides a flat idempotent and commutative forest algebra.

Let $D$ be either a forest diagram or a context diagram over a forest category ${\cal C}.$  Recall that each such diagram has a value $val(D)$ in either 
${\bf HArr}(rootsum(D))$ (for forest diagrams) or ${\bf Arr}(start(D),rootsum(D))$ (for context diagrams).  We denote by $supp(D)$ the set of arrows and half-arrows occurring in the diagram.

\begin{thm}\label{theorem.global_ic}
A forest category is globally idempotent and commutative if and only if the following  condition holds:
If $D_1$ and $D_2$ are forest diagrams over ${\cal C}$ with $supp(D_1)=supp(D_2)$ and $rootsum(D_1)=rootsum(D_2),$ then $val(D_1)=val(D_2).$


\end{thm}
\begin{pr}
First suppose ${\cal C}$ divides a flat idempotent and commutative forest algebra $(H,H).$  Let $D$ be a forest diagram, and let $U$ be the set of all half-arrows and arrows in $D.$  Each $u\in U$ is covered by some $h_u\in H,$ and it follows that $val(D)$ is covered by $\sum_{u\in U} h_u.$  Since $H$ is idempotent and commutative, this value is completely determined by $U.$  Thus if $supp(D_1)=supp(D_2),$ then $val(D_1)$ and $val(D_2)$ are covered by the same element of $H.$ Thus if $rootsum(D_1)=rootsum(D_2)$ as well, $val(D_1)=val(D_2),$ by the injectivity property of division.  
Now suppose ${\cal C}$ satisfies the condition in the statement of the theorem. It follows from faithfulness that ${\cal C}$ satisfies an analogous condition for context diagrams: If $D_1$ and $D_2$ are context diagrams over ${\cal C}$ with $supp(D_1)=supp(D_2),$ $rootsum(D_1)=rootsum(D_2),$ and $start(D_1)=start(D_2),$  then $val(D_1)=val(D_2).$ Let $H$ be the monoid consisting of all subsets of ${\bf HArr}({\cal C})\cup {\bf Arr}({\cal C}),$ with union as the operation.  Suppose $u$ is an arrow or half-arrow of ${\cal C}.$  We say $X\in H$ covers $u$ if there is a context diagram or forest diagram $D$ such that  $val(D)=u$ and $supp(D)=X.$  It follows readily from the conditions on forest diagrams and context diagrams that this covering relation defines a division ${\cal C}\prec (H,H).$

\end{pr}

As corollary to the proof we obtain:

\begin{thm}~\label{theorem.global_ic_decidable}
It is decidable if a given finite forest category is globally idempotent and commutative.
\end{thm}

\begin{pr} The proof of Theorem~\ref{theorem.global_ic} shows that ${\cal C}$ is globally idempotent and commutative if and only if it divides $(H,H),$ where $H$ is the monoid of subsets of ${\bf HArr}({\cal C})\cup {\bf Arr}({\cal C}).$  This can be effectively checked, if necessary by enumerating every possible covering relation and checking if it is a division.

\end{pr}

When one compares the decision procedure given in Theorem~\ref{theorem.global_ic_decidable} for globally idempotent and commutative forest categories to the one in Theorem~\ref{theorem.simon} for ordinary categories, the former appears ridiculously inadequate, and scarcely deserves to be called an `algorithm,' while the latter much more reasonably entails verifying $O(n^2)$ identities, where $n$ is the number of arrows in the category.  Of course one would like something just as reasonable for forest categories. 
More precisely, we wish to have a small list of identities, each involving two diagrams with a small number of objects and arrows, such that one can transform one forest diagram into another with the same support and terminal objects by repeated application of the identities.
It is possible to produce such a list by going carefully through the arguments in
 Place and Segoufin~\cite{ps} on locally testable tree languages, and noting down precisely what axioms are required to produce the analogous result there.  Here is one such list: 
\begin{thm}~\label{theorem.local_global_ic}
Let ${\cal C}$ be a finite forest category in which ${\bf Obj}({\cal C})$ is idempotent and commutative.  ${\cal C}$ is globally idempotent and commutative if and only if
\begin{enumerate}[(i)]
\item (Loop removal.)  Whenever $\stackrel{r}{\rightarrow}y \in{\bf HArr}({\cal C})$ and $y\stackrel{s}{\rightarrow}y,$ $y\stackrel{t_1}{\rightarrow}z_1,$ $y\stackrel{t_2}{\rightarrow}z_2 \in {\bf Arr}({\cal C}),$
$$\begin{array}{c}
\stackrel{r}{\rightarrow}y\stackrel{s}{\rightarrow}y\stackrel{t_1}{\rightarrow}z_1\\
\stackrel{r}{\rightarrow}y\stackrel{s}{\rightarrow}y\stackrel{t_2}{\rightarrow}z_2
\end{array}  =
\begin{array}{c}
\stackrel{r_1}{\rightarrow}y\stackrel{t_1}{\rightarrow}z_1\\
\stackrel{r_2}{\rightarrow}y\stackrel{s}{\rightarrow}y\stackrel{t_2}{\rightarrow}z_2
\end{array}$$
\item (Horizontal absorption) For all $x,y,z,z'\in{\bf Obj}({\cal C}),$ $\stackrel{r}{\rightarrow}x\in{\bf HArr}({\cal C}),$
$\stackrel{s}{\rightarrow}x+y\in{\bf HArr}({\cal C}),$ $t\in{\bf Arr}(x+y,z),$ and $u\in{\bf Arr}(x,z'),$ 

$$\begin{array}{c}\left. \begin{array}{c}\stackrel{r}{\rightarrow}x\\
\stackrel{s}{\rightarrow}x+y\end{array}\right\}\stackrel{t}{\rightarrow}z\\
\stackrel{r}{\rightarrow}x\stackrel{u}{\rightarrow}z'\end{array}
=
\begin{array}{c}
\stackrel{s}{\rightarrow}x+y\stackrel{t}{\rightarrow}z\\
\stackrel{r}{\rightarrow}x\stackrel{u}{\rightarrow}z'
\end{array}
$$
\item (Horizontal idempotence.)  For all $\stackrel{r}{\rightarrow}x\in{\bf HArr}({\cal C}),$
$$\stackrel{r}{\rightarrow}x=\stackrel{r}{\rightarrow}x+\stackrel{r}{\rightarrow}x.$$
\end{enumerate}
 
\end{thm}

The rather involved proof, which naturally enough closely tracks the one given in ~\cite{ps}, is given in the appendix.

    \section{Application to Locally Testable Forest Languages}

In the present section we show how the theory developed in this paper leads to a new treatment of the recent results in ~\cite{ps} on locally testable tree languages.  While our treatment comes wrapped in a great deal of new formalism, it has the advantage of very clearly separating the two main principles of the argument:  The characterization of globally idempotent and commutative forest categories given in Theorems~\ref{theorem.global_ic_decidable} and \ref{theorem.local_global_ic} above, and bounds on the index of definiteness given in Lemma~\ref{lemma.delay_theorem} below.  Each of these principles can be applied separately in other problems.  While our exposition here concerns only what ~\cite{ps} calls `Idempotent Local Testability', we have little doubt that our methods can also shed light on the other formulations of local testability given there.  We discuss these briefly in the final section.

Let $A$ be a finite alphabet and let $s\in H_A.$  Let $k\geq 0.$  We will define the {\it $k$-definite type} of a node in $s$ by induction on $k.$  All nodes have the same 0-definite type.  If $k>0$ then the $k$-definite type of a node is the pair $(a,T),$ where $a$ is the label of the node, and  $T$ is the set of $(k-1)$-definite types of its children.

Let $s,t\in H_A.$  We define $s\sim_k t$ if the set of $k$-definite types of the roots of $s$ is equal to the set of $k$-definite types of the roots of $t.$  It is clear that this equivalence relation is compatible with addition in $H_A,$ and that  $s\sim_k t$ and $p\in V_A$ implies $sp\sim_k tp.$  So $\sim_k$ is a forest algebra congruence of finite index, and thus there is a quotient forest algebra $(H_k,V_k).$  We denote by $\beta_k$ the projection homomorphism from $A^{\Delta}$ onto this quotient.  Note that if $s\in H_A$ then $\beta_k(s)$ can be thought of as the set of $k$-definite types of the root nodes of $s.$ The congruence $\sim_k$ is an analogue for forest algebras to the congruence that identifies two words if they have the same suffix of length $k,$ and the quotient algebra is an analogue to the free $k$-definite semigroup.  In fact, many such analogues are possible, depending on how one defines the horizontal component of the quotient algebra; here we are just treating the case where the horizontal component is idempotent and commutative. 

Again, let $s,t\in H_A,$ and let $k>0.$   We define $s\equiv_k t$ if $s\sim_{k-1} t,$  and the set of $k$-types of nodes in $s$ is equal to the set of $k$-types of nodes of $t.$  Once again, this is a congruence of finite index on $(H_k,V_k).$  We say that $L\subseteq  H_A$ is {\it $k$-locally testable} if it is a union of $\equiv_k$-classes, and {\it locally testable} if it is $k$-locally testable for some $k.$  Thus, for example, membership in a 1-locally testable forest language depends only on the set of node labels for a forest, while a condition like `there is a node labeled $a$ with a child labeled $b,$ but no node labeled $a$ with children labeled $a$ and $b$' defines  a 2-locally testable forest language.

As is the case with words, locally testable languages are recognized by a particular kind of wreath product. The proof of the theorem below is an immediate consequence of the characterization of wreath products in terms of sequential compositions (Theorem 3 of~\cite{bsw}), and the fact that the languages recognized by flat idempotent and commutative algebras are exactly those for which membership only depends on the set of node labels.

\begin{thm}\label{theorem.wreath_lt}
$L\subseteq H_A$ is $k$-locally testable if and only if it is recognized by a homomorphism
$$\gamma:A^{\Delta}\to (H,H)\circ (H_k,V_k),$$
where $(H,H)$ is flat idempotent and commutative, and $\pi\gamma=\beta_k,$ where $\pi$ is the projection homomorphism from the wreath product onto its right-hand factor, 
\end{thm}

The following lemma, a critical combinatorial fact in this study, is adapted from another argument in Place and Segoufin~\cite{ps}.  In many respects, it plays the role of the `Delay Theorem' (Tilson~\cite{tilson}) in analogous work on languages of words.

\begin{lemma}~\label{lemma.delay_theorem}
Let $(H,V)$ be a finite forest algebra, with $H$ idempotent and commutative, and let $\alpha:A^{\Delta}\to (H,V)$ be a homomorphism.  Let $k>|H|^2,$ and let $N>k.$  Then the following properties hold:

\begin{enumerate}[(i)]
\item If $r,s\in H_A$ with $\beta_k(r )+\beta_k(s)=\beta_k(s),$ then there exist $r ',s'\in H_A$ such that $\alpha(r )=\alpha(r'),$ $\alpha(s)=\alpha(s'),$ and $\beta_N(r')+\beta_N(s')=\beta_N(s').$
\item If $r\in H_A,$ $p\in V_A,$ with $\beta_k(r)=\beta_k(rp),$ then there exist $r'\in H_A,$ $p'\in V_A,$ with $\alpha(r)=\alpha(r'),$ $\alpha(rp)=\alpha(r'p'),$ and $\beta_N(r')=\beta_N(r'p').$
\end{enumerate}
\end{lemma}

 The proof is given in the appendix.

For local testability, this implies the following:

\begin{thm}~\label{theorem.lt_bound}
If $L\subseteq H_A$ is locally testable, then it  is $|H|^2+1$-locally testable.
\end{thm}

\begin{pr}
By hypothesis, $L$ is $N$-locally testable for some $N.$  Let $k=|H_L|^2+1.$   By Theorems~\ref{theorem.wreath_lt} and \ref{theorem.derived_category}, ${\cal D}_{\alpha_L,\beta_N}$ divides a flat idempotent and commutative forest algebra.  This implies, by Theorem~\ref{theorem.local_global_ic}, that ${\bf HArr}({\cal D}_{\alpha_L,\beta_N})$ is idempotent and commutative, and consequently its homomorphic image $H_L$ is idempotent and commutative.  Thus Lemma~\ref{lemma.delay_theorem} applies.  We will use this lemma to show that ${\cal D}_{\alpha_L,\beta_{k}}$ satisfies the three conditions in Theorem~\ref{theorem.local_global_ic}, and thus $L$ is $(k+1)$-locally testable by Theorems~\ref{theorem.wreath_lt}.  We have already observed that $H_L$ is idempotent and commutative, and this gives us horizontal idempotence of ${\cal D}_{\alpha_L,\beta_{k}}$ for all values of $k.$

To establish the horizontal absorption condition, let $r,s\in H_A$ with $\beta_{k}(r)+\beta_{k}(s)=\beta_{k}(s).$  We need to show, for all $t,u\in V_A,$ that $\alpha_L((r+s)t+ru)=\alpha_L(st+ru)$---this is precisely what it means for the two half-arrows on the two sides of  the horizontal absorption identity to be equal in the derived category. If we take $r'$ and $s'$ as in the Lemma, then since the horizontal absorption identity is assumed to hold in ${\cal D}_{\alpha_L,\beta_N},$ we have $\alpha_L((r'+s')t+r'u)=\alpha_L(s't+r'u).$ Since $\alpha_L(r)=\alpha_L(r')$ and $\alpha_L(s)=\alpha_L(s'),$ we obtain $\alpha_L((r+s)t+ru)=\alpha_L(st+ru),$ as required.  The loop removal condition is established in the same way, using the other part of Lemma~\ref{lemma.delay_theorem}.
\end{pr}

\begin{thm}\label{theorem.lt_decidable}
It is decidable whether a given regular forest language $L$ (given, say, by an automaton that recognizes it) is locally testable.
\end{thm}

\begin{pr}
Theorems~\ref{theorem.lt_bound} and Theorem~\ref{theorem.global_ic_decidable} give an easy proof of decidability:  From the presentation of $L$ we can effectively calculate the syntactic forest algebra $(H_L,V_L),$ the syntactic morphism $\alpha_L,$ and from this the derived category ${\cal D}_{\alpha,\beta_{k}},$ where $k$ is as in Theorem~\ref{theorem.lt_bound}. We can then check effectively whether this category is globally idempotent and commutative.

\end{pr}

We get a nicer proof of decidability by employing the criteria in Theorem~\ref{theorem.local_global_ic}. As we saw in the proof of Theorem~\ref{theorem.lt_bound}, the category conditions translate into simple identities in the same spirit as the `tameness' conditions given in ~\cite{ps}. We state these formally as follows:

\begin{thm} $L\subseteq H_A$ is locally testable if and only if the following hold, with $k={|H|}^2+1.$ 
\begin{enumerate}[(i)]
\item For all $r,s\in H_A,$ $t,u\in V_A,$ with $\beta_{k}(r)\subseteq\beta_{k}(s),$ we have
$$\alpha_L((r+s)t+ru)=\alpha_L(st+ru).$$  

\item For all $r\in H_A,$ $p,q,q'\in V_A,$ with $\beta_k(rp)=\beta_k(r),$ we have
$$\alpha_L(rpq+rpq')=\alpha_L(rq+rpq').$$

\end{enumerate}
\end{thm}

\section{Conclusion and Further Research}
We have shown here how to extend the algebra of finite categories, which Tilson~\cite {tilson} described as `an essential ingredient' in the study of monoids---and especially the Derived Category Theorem---to the setting of forest algebras.  This entailed some fundamental modifications to the original definition, in particular the introduction of an additive structure on ${\bf Obj}({\cal C})$ and the use of half-arrows.  We then showed how this can be applied to give a new treatment recent results on locally testable languages.

The great advantage of this abstract approach is that, as with word languages and monoids, it isolates the mathematical principles underlying the separate parts of the argument in such a manner that they can be applied elsewhere to a range of other problems.  Let us briefly indicate what some of these other problems might be.

First, we chose the case of idempotent locally testable languages because they were in a sense the easiest to treat. But Place and Segoufin treat another version of local testability, called $(k,l)$-local testability, in which one looks not merely for occurrences of neighborhoods of depth $k,$ but counts the number of these occurrences up to threshold $l.$ Here again the problem becomes one of determining whether the syntactic morphism factors through a wreath product with a flat idempotent and commutative forest algebra on the left. If one can bound $k$ and $l$ in terms of the size of the syntactic forest algebra (carried out in ~\cite{ps} by a pumping argument along the lines of our Lemma~\ref{lemma.delay_theorem}), then our Theorem~\ref{theorem.global_ic_decidable} applies to give decidability.  Unfortunately, we no longer have the more satisfactory Theorem~\ref{theorem.local_global_ic} since in this case the object set of the derived category is no longer idempotent.  Finding a version of this theorem that works without the hypothesis of an idempotent object set is an worthwhile problem.  It would also be worthwhile to see how to make this theory work for binary trees, the first case Place and Segoufin treat. These do not fit so neatly into our formalism, which was developed for unranked trees.

One can also try to apply these methods to the treatment of tree languages definable in first-order logic with successor, in ~\cite{ben-seg}, where a wreath product with a flat aperiodic and commutative left-hand factor occurs.

The large number of problems in the theory of languages and monoids that involved wreath products where the right-hand factors are definite semigroups led to the formulation of a general principle, by means of which one could bound the index of definiteness as a function of the size of the syntactic monoid. This is the `Delay Theorem', established first in different terms in Straubing~\cite{v*d} and given its definitive formulation in terms of categories by Tilson~\cite{tilson}.  In our application of forest algebras, we bound this index only in our very special case (Lemma~\ref{lemma.delay_theorem}), but we strongly suspect that the same sort of argument can be extended to give a more general formulation, an analogue of the Delay Theorem for forest categories.

Finally, we have already noted that several important unsolved problems about logics on trees ({\it e.g.,} $CTL,$ $CTL^*,$ and first-order logic with ancestor) hinge on being able to decide whether a given forest algebra admits a certain kind of wreath product decomposition~\cite{bsw}. The Derived Category Theorem, both in its original formulation and in the extension we give here to forest algebras, was designed precisely to address the question of finding such decompositions. So it may well prove to play an important role in the solution to these problems from logic.
\bibliographystyle{plain}
\bibliography{forest_categories2}


\appendix
\section{Proof of the Derived Category Theorem}

Before proceeding to the proof, we note a subtlety in the definition of division of forest algebras. Since a forest algebra is, in particular, a transformation monoid, there is actually a second notion of division, which comes from the theory of transformation monoids:  We say that  $(H,V)$ {\it tm-divides} $(H',V')$ if there is a submonoid $K$ of $H',$ and a surjective monoid homomorphism $\Psi:K\to H$ such that for each $v\in V$ there exists $\hat v \in V'$ with $K\hat v\subseteq K,$ and for all $k\in K,$
$$\Psi(k\hat v)=\Psi(k)v.$$
Fortunately, the two notions of division coincide.  It is not difficult to show:

\begin{lemma}\label{lemma.division}
Let $(H_1,V_1)$ and $(H_2,V_2)$ be forest algebras.  $(H_1,V_1)\prec(H_2,V_2)$ if and only if $(H_1,V_1)$ {\it tm}-divides $(H_2,V_2).$
\end{lemma}
\begin{pr}
 First suppose $(H_1,V_1)$ divides $(H_2,V_2).$  Then there is a submonoid
$V'$ of $V_2$ and a  forest algebra homomorphism 
$$\alpha:(0\cdot V',V')\to (H_1,V_1).$$
(Strictly speaking, we should reduce $V'$ to the quotient that acts faithfully on $0\cdot V',$ but leaving this reduction out does not change the argument.)  Let $v\in V_1,$ and set $\hat v$ to be any 
element of $V'$ such that $\beta({\hat v})=v.$  We then have for $h\in 0\cdot V',$

$$\alpha(h{\hat v})=\alpha(h)\alpha({\hat v})=\alpha(h)v,$$
so $(H_1,V_1)$ tm-divides $(H_2,V_2).$ 



Conversely, suppose  $(H_1,V_1)$ tm-divides $(H_2,V_2),$ with underlying morphism $\alpha:H'\to H_1.$   Let $A$ be an alphabet at least as large as $V_1,$ and let $\gamma:A\to V_1$
be an onto map.  This extends, because of the universal property of the free forest algebra, to a (surjective) forest algebra morphism $\gamma:A^{\Delta}\to(H_1,V_1).$  We define $\delta:A\to V_1'$ by setting
$$\delta(a)=\widehat{\gamma(a)}$$
for all $a\in A,$ and consider its  extension $\delta$ to a forest algebra morphism.
It is enough to show that for $x,y\in V_{A},$ $\delta(x)=\delta(y)$ implies $\gamma(x)=\gamma(y).$  This will imply that $\gamma$ factors through $\delta.$ and give the required division.

Observe that if $s\in H_{A},$ then $\delta(s)$ is in the domain $H'$ of $\alpha,$ because $s=0\cdot x$ for some $x\in V_1,$ and thus  
\begin{eqnarray*}
\gamma(s) &=&\gamma(0)\gamma(x)\\
                    &=&\alpha(\delta(0))\gamma(x)\\
                    &=&\alpha(\delta(0)\widehat{\gamma(x)})\\
                    &=&\alpha(\delta(0)\delta(x))\\
                    &=&\alpha(\delta(0\cdot x))\\
                    &=&\alpha(\delta(s)).
\end{eqnarray*} 
So by assumption, we have
$$\alpha(\delta(s))\gamma(a)=\alpha(\delta(s)\delta (a))$$
for all $s\in H_{A},$ $a\in A$  A straightforward induction on the number of nodes in $x$ implies that for any $x\in V_{A},$
$$\alpha(\delta(s))\gamma(x)=\alpha(\delta(s)\delta(x)).$$
Now suppose $h\in H_1$ and $\delta(x)=\delta(y). $  As noted above, $h=\alpha(\delta(s))$ for some
$s\in H_{A},$ and consequently
\begin{eqnarray*}
h\cdot\gamma(x) &=&\alpha(\delta(s))\gamma(x)\\
	&=&\alpha(\delta(s)\delta(x))\\
	&=&\alpha(\delta(s)\delta(y))\\
	&=&\alpha(\delta(s))\gamma(y)\\
	&=&h\cdot\gamma(y).
\end{eqnarray*}

Since $h$ was arbitrary, we get $\gamma(x)=\gamma(y),$ by faithfulness.

\end{pr}

We now proceed to the proof of Theorem~\ref{theorem.derived_category}. We first prove part {\it (a)}.  By Lemma~\ref{lemma.division}
we need to exhibit a surjective partial function
$$\Psi:H\times H_1\to H_1$$
such that ${\rm dom}\Psi $ is a submonoid of $H\times H_2$ and $\Psi\big\vert_{{\rm dom}\Psi}$ is a homomorphism, and to define for each $v\in V_1$ an element $\hat v$ of the vertical monoid of $(H,V)\circ (H_2,V_2)$ such that for all $(h,h_2)\in{\rm dom}\Psi,$ $(h,h_2)\hat v\in {\rm dom}\Psi,$ and 
$$\Psi((h,h_2)\hat v)=\Psi(h,h_2)v.$$

To this end, we define
$\Psi(h,h_2)=h_1$
if $\stackrel{h_1}{\rightarrow} h_2$ is a half-arrow covered by $h.$  This is surjective, since every $h_1\in H_1$ is $\alpha(s)$ for some $s\in H_A.$  We need to verify that $\Psi$ is well-defined; that is we cannot have $h_1\neq h_1'$ with $h_1'=\Psi(h,h_2)=h_1.$  This follows from the injectivity property of division:  Two different half-arrows with the same end cannot be covered by the same element of $H.$  Also, from the definition of division, if $h,h'$ cover $\stackrel{h_1}{\rightarrow} h_2,$ and $\stackrel{h_1'}{\rightarrow} h_2,'$ respectively, then $h+h'$ covers the sum $\stackrel{h_1+h_1'}{\longrightarrow} h_2+h_2',$ so that $(h+h',h_2+h_2')\in {\rm dom}\Psi,$ and
$\Psi(h+h',h_2+h_2')=h_1+h_1'.$  Thus ${\rm dom}\Psi$ is a submonoid of $H\times H_1,$ and  the restriction of $\Psi$ to its domain is a homomorphism.

 Now let $v\in V_1.$  Then $v=\alpha(p)$ for some $p\in V_A.$  For each $h_2\in H_2,$ define $f_p(h_2)$ to be any element of $V$ that covers $h_2\stackrel{p}{\rightarrow}h_2\cdot\beta(p).$  Set $\hat v=(f_p,\beta(p)).$  Now suppose $\Psi(h,h_2)=h_1.$  Then 
 $$(h,h_2)\hat v=(h\cdot f_p(h_2),h_2\cdot\beta(p)).$$
 By the defininiton of division,
 \begin{eqnarray*}
 h\cdot f_p(h_2)&\in& K_{\stackrel{h_1}{\rightarrow}h_2}\cdot K_{h_2\stackrel{p}{\rightarrow}h_2\cdot\beta(p)}\\
 &\subseteq & K_{\stackrel{h_1\cdot\alpha(p)}{
 \longrightarrow}h_2\beta(p)},\\
 \end{eqnarray*}
 so $(h,h_2)\hat v\in{\rm dom}\Psi$ and
 \begin{eqnarray*}
 \Psi((h,h_2)\hat v)&=&h_1\cdot\alpha(p)\\
 &=& h_1v.\\
 \end{eqnarray*}
 
 We now prove part {\it (b)}.  For $s\in H_A,$ $p\in V_A,$ we write $h_s$ for the left component of $\delta(s),$ and $f_p$ for the left component of $\delta(p).$  We set
 $$K_{\stackrel{h_1}{\rightarrow}h_2}=\{h_s:s\in H_A,\alpha(s)=h_1,\beta(s)=h_2\},$$
 $$K_{h\stackrel{p}{\rightarrow}h'}=\{f_q(h): q\in V_A, (h,q,h')\sim (h,p,h')\}.$$
 
 We need to show that these covering relations define a division, so we have to verify both the operation-preserving and injectivity properties.
 
 For injectivity, suppose first that $h\in H$ covers both $\stackrel{h_1}{\rightarrow}h_2,$ and $\stackrel{h_1'}{\rightarrow}h_2.$  Then there exist $s,t\in H_A$ such that $\alpha(s)=h_1,$ $\alpha(t)=h_1',$ $\beta(s)=\beta(t)=h_2,$ and $h_s=h_t.$  Thus $\delta(s)=\delta(t),$ so 
 $$h_1=\alpha(s)=\gamma\delta(s)=\gamma\delta(t)=h_1'.$$
 Now suppose that $v\in V$ covers both $h\stackrel{p}{\rightarrow}h'$ and $h\stackrel{p'}{\rightarrow}h'.$  Then there exist $q,q'\in V_A$ such that 
 \begin{eqnarray*}
 h\stackrel{p}{\rightarrow}h' &=& h\stackrel{q}{\rightarrow}h'\\
 h\stackrel{p'}{\rightarrow}h' &=& h\stackrel{q'}{\rightarrow}h' \\
 f_q(h)&=&f_{q'}(h).
  \end{eqnarray*}
  Let $\beta(s)=h.$ Then
  \begin{eqnarray*}
\delta(sq) &=& \delta(s)\delta(q)\\
&=&(h_s,\beta(s))(f_q,\beta(q))\\
&=&(h_s,h)(f_q,\beta(q))\\
&=&(h_s\cdot f_q(h),h\cdot\beta(q))\\
&=&(h_s\cdot f_q(h),h'),\\
\end{eqnarray*}
and likewise $\delta(sq')=(h_sf_{q'}(h),h'),$ and thus $\delta(sq)=\delta(sq'),$ so $\alpha(sq)=\alpha(sq'),$ and consequently $h\stackrel{p}{\rightarrow}h'= h\stackrel{p'}{\rightarrow}h'.$  This proves injectivity.

We now verify the operation-preserving properties of division.  First, if $h, h'\in H$ cover $\stackrel{h_1}{\rightarrow} h_2,$ $\stackrel{h_1'}{\rightarrow} h_2',$ respectively, then there exist $s,s'\in H_A$ such that
$$\delta(s)=(h,h_2),$$
$$\delta(s')=(h',h_2),$$
and $\alpha(s)=h_1,$ $\alpha(s')=h_1'.$  It follows immediately that $h+h'$ covers
$$\stackrel{h_1+h_1'}{\longrightarrow} h_2+h_2'\quad=\quad \stackrel{h_1}{\rightarrow} h_2+\stackrel{h_1'}{\rightarrow} h_2'.$$

Next, if $v_1,v_2\in V$ cover $h\stackrel{p_1}{\rightarrow} h',$ $h'\stackrel{p_2}{\rightarrow} h'',$ respectively, then there exist $q_1,q_2\in V_A$ such that $v_1=f_{q_1}(h),$ $v_2=f_{q_2}(h')=f_{q_2}(h\cdot \beta(q_1)).$  Now, from the definition of the wreath product we have

\begin{eqnarray*}
\delta(q_1q_2)&=&\delta(q_1)\delta(q_2)\\
&=&(f_{q_1},\beta(q_1))\cdot (f_{q_2},\beta(q_2))\\
&=&(f_{q_1q_2},\beta(q_1q_2)),
\end{eqnarray*}
where for $h\in H,$ $f_{q_1q_2}(h)=f_{q_1}(h)\cdot f_{q_2}(h\cdot\beta(q_1)).$  Thus $v_1v_2=f_{q_1q_2}(h),$ and so $v_1v_2$ covers $h\stackrel{p_1}{\rightarrow} h'\stackrel{p_2}{\rightarrow} h''.$ If, further, $k\in H$ covers $\stackrel{h_1}{\rightarrow}h,$ then for some $s\in H_A,$ $\delta(s)=(k,h).$  Thus
\begin{eqnarray*}
\delta(sq_1)&=&(k,h)(f_{q_1},\beta(q_1))\\
&=&(k\cdot f_{q_1}(h),h\cdot\beta(q_1))\\
&=&(kv_1,h'),\\
\end{eqnarray*}
so that $kv_1$ covers

\begin{eqnarray*}
\stackrel{\alpha(sq_1)}{\longrightarrow}h'&=&\stackrel{h_1}{\rightarrow}h\stackrel{q_1}{\rightarrow}h'\\
&=&\stackrel{h_1}{\rightarrow}h\stackrel{p_1}{\rightarrow}h'.
\end{eqnarray*}

Finally, observe that for $q\in V_A,$ $s\in H_A,$ and $h\in H_2,$
$$f_{q+s}(h_2)=f_q(h_2)+h_s,$$
and it follows readily that if $v$ covers $h_2\stackrel{q}{\rightarrow}h_2'$ and $h_s$ covers
$\stackrel{h_1}\rightarrow h_2'',$ then $v+h_s$ covers $h_2\stackrel{q}{\rightarrow}h_2'+\stackrel{h_1}\rightarrow h_2''.$

\section{Proof of Theorem~\ref{theorem.local_global_ic}}

One direction of the theorem is a trivial consequence of Theorem ~\ref{theorem.global_ic}:  If ${\bf Obj}({\cal C})$ is idempotent and commutative, then each of the three identities is a pair of forest diagrams with the same support and the same rootsum. So if the category is globally idempotent and commutative, the two diagrams have the same value in ${\cal C},$ and thus the identity is satisfied. 

Conversely,  suppose ${\bf Obj}({\cal C})$ is idempotent and commutative, and that the three identities are satisfied.  We will begin by deducing several additional identities from this initial list of three axioms.

\noindent {\it (i)} {\it (Vertical idempotence)}  For all arrows $u\in {\bf Arr}(x,x),$ $uu=u.$  To see this, let $u=x\stackrel{s}{\rightarrow}x,$ and let $\stackrel{r}{\rightarrow}x\in{\bf HArr}({\cal C}).$  We have, by several applications of horizontal idempotence and loop removal,
$$
\begin{array}{l}
 \stackrel {r}{\rightarrow}x\stackrel{s}{\rightarrow}x\stackrel{s}{\rightarrow}x =\\
\stackrel {r}{\rightarrow}x\stackrel{s}{\rightarrow}x\stackrel{s}{\rightarrow}x+\stackrel {r}{\rightarrow}x\stackrel{s}{\rightarrow}x\stackrel{s}{\rightarrow}x =\\
\stackrel {r}{\rightarrow}x\stackrel{s}{\rightarrow}x+\stackrel {r}{\rightarrow}x\stackrel{s}{\rightarrow}x\stackrel{s}{\rightarrow}x =\\
\stackrel {r}{\rightarrow}x\stackrel{s}{\rightarrow}x+\stackrel {r}{\rightarrow}x\stackrel{s}{\rightarrow}x = \\
 \stackrel {r}{\rightarrow}x\stackrel{s}{\rightarrow}x.
 \end{array}.
 $$
 Since this holds for all $\stackrel{r}{\rightarrow}x\in{\bf HArr}(x),$ by faithfulness we have $x\stackrel{s}{\rightarrow}x\stackrel{s}{\rightarrow}x= x\stackrel{s}{\rightarrow}x.$ 
 \medskip
 
 \noindent {\it (ii)} {\it (Horizontal swap)}  For any $r,s\in {\bf HArr}(x),$ $t\in{\bf Arr}(x,y),$ $u\in{\bf Arr}(x,z),$
 $$\stackrel{r}{\rightarrow}x\stackrel{t}{\rightarrow}y+\stackrel{s}{\rightarrow}x\stackrel{u}{\rightarrow}z=\stackrel{s}{\rightarrow}x\stackrel{t}{\rightarrow}y+\stackrel{r}{\rightarrow}x\stackrel{u}{\rightarrow}z.$$
 We have by repeated use of horizontal absorption,
 $$
 \begin{array}{l}
 \stackrel{r}{\rightarrow}x\stackrel{t}{\rightarrow}y+\stackrel{s}{\rightarrow}x\stackrel{u}{\rightarrow}z =\\
 ( \stackrel{r}{\rightarrow}x+\stackrel{s}{\rightarrow}x)\stackrel{t}{\rightarrow}y+\stackrel{s}{\rightarrow}x\stackrel{u}{\rightarrow}z =\\
 \stackrel{r}{\rightarrow}x\stackrel{1_X+s}{\rightarrow}x\stackrel{t}{\rightarrow}y+\stackrel{s}{\rightarrow}x\stackrel{u}{\rightarrow}z =\\
   \stackrel{r}{\rightarrow}x\stackrel{1_X+s}{\rightarrow}x\stackrel{t}{\rightarrow}y+(\stackrel{s}{\rightarrow}x+\stackrel{r}{\rightarrow}x) \stackrel{u}{\rightarrow}z =\\
   ( \stackrel{r}{\rightarrow}x+\stackrel{s}{\rightarrow}x)\stackrel{t}{\rightarrow}y+ (\stackrel{s}{\rightarrow}x+\stackrel{r}{\rightarrow}x) \stackrel{u}{\rightarrow}z.     \end{array}
 $$
 By symmetry, this is also equal to 
 $$\stackrel{s}{\rightarrow}x\stackrel{t}{\rightarrow}y+\stackrel{r}{\rightarrow}x\stackrel{u}{\rightarrow}z,$$ 
 giving the required result.
 
 \noindent{\it (iii)}  For any $\stackrel{r}{\rightarrow} x,$ $\stackrel{t}{\rightarrow}y\in {\bf HArr}({\cal C}),$ and any $x+y\stackrel{s}{\rightarrow} z,$ $x\stackrel{u}{\rightarrow}x+y\in{\bf Arr}({\cal C}),$ we have
 
  $$\begin{array}{l}(\stackrel{r}{\rightarrow} x + \stackrel{t}{\rightarrow}y)\stackrel{s}{\rightarrow}z+\stackrel{r}{\rightarrow}x\stackrel{u}{\rightarrow}x+y=\\
  (\stackrel{r}{\rightarrow}x\stackrel{u}{\rightarrow}x+y\quad+\quad\stackrel{t}{\rightarrow}y)\stackrel{s}{\rightarrow}z+\stackrel{r}{\rightarrow}x\stackrel{u}{\rightarrow}x+y.
  \end{array}$$
  To see this, note that by horizontal absorption,
  {\small
  $$
  \begin{array}{l}
  (\stackrel{r}{\rightarrow} x + \stackrel{t}{\rightarrow}y)\stackrel{s}{\rightarrow}z+\stackrel{r}{\rightarrow}x\stackrel{u}{\rightarrow}x+y  =\\
  (\stackrel{r}{\rightarrow} x + \stackrel{t}{\rightarrow}y+\stackrel{r}{\rightarrow}x\stackrel{u}{\rightarrow}x+y  )\stackrel{s}{\rightarrow}z+\stackrel{r}{\rightarrow}x\stackrel{u}{\rightarrow}x+y = \\
  (\stackrel{r}{\rightarrow}x\stackrel{u}{\rightarrow}x+y\quad+\quad\stackrel{t}{\rightarrow}y)\stackrel{s}{\rightarrow}z+\stackrel{r}{\rightarrow}x\stackrel{u}{\rightarrow}x+y. 
  \end{array}
  $$ }
  
  \noindent{\it (iv)}{\it (Horizontal transfer)}   Here the setup is a little different.  We imagine a {\it multicontext diagram} $D$.  This is like a context diagram except that we allow more than one object to be exposed at the leaves.  If these objects are $x_1,x_2,x_3$ in some order, then we can view $D$ as defining a function from triples of half-arrows with ends $x_1,x_2,x_3$ to half-arrows with end $rootsum(D).$  We denote the value of this function at half-arrows $u_1,u_2,u_3$ by $D(u_1,u_2,u_3).$  Now let us suppose that the three objects are $x+y,x$ and $x+y$ respectively, and the three half-arrows are $v+w,$ $v$ and $u,$ where $end(u)=x+y,$ $end(v)=x,$ and $end(w)=y.$  Then
  $$D(v+w,v,u)=D(u+w,v,u).$$
  As in the preceding examples, this is proved by several successive applications of horizontal absorption.
  
  We will use the above properties to establish the following Lemma.
  
  \begin{lemma}\label{lemma.switcheroony}
  Let ${\cal C}$ be a forest category in which ${\bf Obj}({\cal C})$ is idempotent and commutative, and that satisfies the three conditions in the hypothesis of Theorem~\ref{theorem.local_global_ic}.  Suppose that $D$ is a forest diagram such that $D=D_1E_1=D_2E_2,$ where $D_1$ and $D_2$ are forest diagrams that do not overlap, and $E_1,E_2$ are context diagrams such that $supp(D_1)\subseteq supp(E_1),$ and $rootsum(D_1)=rootsum(D_2).$  Then  $val(D)=val(D_2E_1).$
  \end{lemma}
  
  Assuming for now the truth of Lemma~\ref{lemma.switcheroony}, we complete the proof of Theorem~\ref{theorem.local_global_ic}.  Let $D_1,D_2$ be forest diagrams with $rootsum(D_1)=rootsum(D_2)$ and $supp(D_1)=supp(D_2).$  We will show $val(D_1)=val(D_1+D_2).$  By symmetry, we will also get $val(D_2)=val(D_1+D_2),$ so
 $val(D_1)=val(D_2),$ as required.  To establish $val(D_1)=val(D_1+D_2),$ we argue by induction on the number of tree components of $D_2.$  We write $D_2=D_2'+T,$ where $T$ is a tree diagram, and we will prove $val(D_1+D_2)=val(D_1+D_2').$  In this manner we will eliminate every tree occurring in $D_2$ and eventually get the desired result.
 
 $T$ consists of either a single half-arrow $u,$ or has the form $D_3u,$ where $u$ is an arrow.  In the former case, the same half-arrow occurs at a leaf in $D_1,$ because $supp(D_1)=supp(D_2).$  Thus $D_1+D_2=uE+u$ for some context $E$ such that $rootsum(E)+rootsum(u)=rootsum(E).$  Horizontal absorption implies $val(D_1+D_2)=val(uE) = val(D_1+D_2').$  In the latter case, where $u$ is an arrow, the same arrow occurs somewhere in $D_1.$  Let $S$ be the tree diagram whose root is this occurrence of $u.$  We can write $D_1+D_2=T+SE_2,$ where $supp(T)\subseteq supp(SE_2)$ and $rootsum(T)=rootsum(S).$  Thus by Lemma~\ref{lemma.switcheroony}, $val(D_1+D_2)=val(S+SE_2),$ and by horizontal absorption, this is $val(SE_2)=val(D_1+D_2').$
 
 We now turn to the proof of the Lemma itself. We assume $D,D_1,D_2,E_1,E_2$ are as in the statement of the Lemma.  We argue by induction on the depth of $D_1.$  The base step is when $D_1$ has depth 1, so that $D_1=u_1+\cdots+u_r,$ where the $u_i$ are all half-arrows.  We will show how to replace each $u_i$ in turn by $D_2$ without changing the value of the diagram, so that in the end $val(D)=val((D_2+\cdots +D_2)E_1)=val(D_2E_1),$ by horizontal idempotence. Suppose $i\geq 1,$ and that we have already replaced the $u_j$ with $j<i.$  Since $supp(D_1)\subseteq supp(E_1),$ the half-arrow $u_i$ occurs somewhere outside of $D_1.$  If this occurrence is outside of $D_2$ as well, then we can write
 {\small $$val(D)=val(E(D_2+\cdots+D_2+u_i+\cdots+u_r,u_i,D_2)),$$}
 where $E$ is a multicontext.
 (Note that if $i=1,$ there is no occurrence of $D_2$ in the leftmost argument to $E.$  Our reasoning needs to work in this case as well.)
 By horizontal transfer, this is equal to  $val(E(D_2+\cdots+D_2+u_{i+1}+\cdots+u_r,u_i,D_2)).$ If, on the other hand, the occurrence of $u_i$ is within $D_2,$ then we write
{\small $$\begin{array}{l}val(D)=\\val((D_2+\cdots +D_2+u_i+u_{i+1}+\cdots u_r)F_1+\\u_iF_2G_2),\end{array}$$}
where $u_iF_2=D_2.$  By property {\it (iii)} above, we can replace the first $u_i$ by $u_iF_2=D_2,$ and thus eliminate $u_i.$

We now proceed to the inductive step, so that $D_1$ has depth greater than 1.  Thus $D_1$ is a sum of tree diagrams $D_1=T_1+\cdots + T_r,$ where at least one of the $T_i$ has depth greater than 1.  Once again, we will show how to replace the $T_i,$ one at a time, by copies of $D_2.$  We already know how to do this when $T_i$ is a half-arrow, so we assume $T_i$ has depth greater than 1.  We thus write $T_i=Fv,$ where $F$ is a forest diagram and $v$ is an arrow.  By assumption, the arrow $v$ occurs somewhere in $E_1.$  Thus there are two cases to consider: The occurrence of  $v$ in $E_1$ can be an ancestor of $T_i,$ or not.  If it is not an ancestor, then we can write
$$D=(FvF'+GvG')H.$$
Now every half-arrow and arrow in $F$ also appears outside of $F,$ and $rootsum(F)=rootsum(G)=start(v).$  Since the depth of $F$ is strictly less than the depth of $D_1,$ we can apply the inductive hypothesis and replace $F$ by $G,$ so that
$$val(D)=val((GvF'+GvG')H).$$
From this point, the proof is identical to the argument given by Place and Segoufin~\cite{ps}:  There are three possibilities, depending where $D_2$ appears in relation to the right occurrence of $Gv$ in the above expression:  It could occur either inside $Gv$, or contain $Gv,$ or not overlap $Gv.$  In all three cases we can use our axioms and their consequences to replace the left occurrence of $Gv$ by $D_2,$ with the result that $T_i$ has been replaced by $D_2.$  This leaves us with the case where the occurrence of the arrow $v$ in $E_1$ is an ancestor of the root of $T_i.$  We distinguish two subcases, depending on whether or not the original occurrence of $D_2$ in $D$ is a descendant of $v.$  We use horizontal swap in one case, and vertical idempotence in the other, to reduce to the prior cases.    
  \section{Proof of Lemma~\ref{lemma.delay_theorem}}
  
 We first prove part (i).  Note that the condition $\beta_k(r)+\beta_k(s)=\beta_k(s)$ translates simply to $\beta_k(r)\subseteq\beta_k(s).$     Given a node $x$ of depth $m$ in a forest $r,$  we denote by $trace_r(x)$ the sequence
 $$(a,\kappa_1,\kappa_1,\cdots,\kappa_m),$$
 where $a$ is the label of $x,$ $\kappa_1$ is the 2-definite type of the parent of $x,$ $\kappa_3$ is the 3-definite type of the grandparent of $x,$ {\it etc.}  Finally $\kappa_m$ is the $m$-definite type of the root of $r$ that $x$ is descended from.  We shall also call such a sequence an $(m+1)$-trace of $r.$
 
 Take $r,s,k,$ as in the hypothesis of the Lemma.  Let $d$ be the smallest integer for with there is a $(d+1)$-trace of $r$ that is not in $s.$   We know $d>k,$ because by hypothesis $\beta_k(r)\subseteq\beta_k(s).$ Let $x$ be a node of depth $d$ in $r$ such that $trace_r(x)$ is not a $(d+1)$-trace of $s.$  Let $y$ be the parent of $x.$  By the minimality condition, there exists a node $z$ in $s$ such that $trace_r(y)=trace_s(z).$
 
 We now evaluate the homomorphism $\alpha$ on both $r$ and $s.$  Computation of the value of $\alpha$ on a forest proceeds in a bottom-up fashion from the leaves, attaching a value $h_t\in H$ to each node $t$ of the forest; the final value is then just the sum in $H$ of the values attached to the roots.  Let us look at the two sequences of values
 $$h_y=h_{y_1}, h_{y_2},\ldots, h_{y_d}$$
 $$h_z=h_{z_1}, h_{z_2},\ldots, h_{z_d}$$ 
 attached to the successive nodes along the paths from $y$ and $z$ to the roots.   Since $d>|H|^2,$ there is a pair of indices $i<j$ such that
 $$h_{y_i}=h_{y_j},$$
 $$h_{z_i}=h_{z_j}.$$
 We can then write
 $$r={\bar r}p_1q_1,$$
 $$s={\bar s}p_2q_2,$$
 where ${\bar r}, {\bar s}$ are the trees rooted at $y_i, z_i,$ respectively, and ${\bar r}p_1,$ ${\bar s}p_2\in V_A$ are the trees rooted at $y_j$ and $z_j.$  We now `pump' the contexts $p_1$ and $p_2$ and set 
 $$r_1={\bar r}p_1^2q_1,$$
 $$s_1={\bar s}p_2^2q_2.$$
 From the way in which the contexts $p_1,p_2$ were chosen, we have $\alpha(r_1)=\alpha(r),$ $\alpha(s_1)=\alpha(s).$  Observe that as a result of the pumping we have eliminated a $(d+1)$-trace from $r$ that did not appear in $s,$ and introduced some new $(d+1)$-traces into both $r_1$ and $s_1.$ But---and this is crucial---in passing from $r$ to $r_1,$ we have not introduced any additional $(d+1)$-traces that are not also in $s_1.$  We can thus repeat this process with any $(d+1)$-trace that remains in $r_1$ but is not in $s_1,$ and the number of these will decrease at each step until we have forests $r'',s''$ with $\alpha(r'')=\alpha(r),$ $\alpha(s'')=\alpha(s),$ and $\beta_{d+1}(r'')\subseteq\beta_{d+1}(s'').$  We can now repeat the whole argument with $k$ replaced by $d+1,$ to make the subscript on $\beta$ as large as we like.
 
 We now turn to the proof of the second part.  The proof is very much along the same lines, but we need to attend to the possibility that the paths underlying the two traces overlap. We suppose then $\beta_k(r)=\beta_k(rp),$ and let $d$ denote the minimum depth of a node $s$ such that the trace of $x$ in $r$ does not appear in $rp,$ or vice-versa.  We will suppose that $trace_r(x)$ does not appear in $rp,$ but the argument is the same in the other case.  Again we let $y$ be the parent of $x$ and we find a node $z$ in $rp$ such that $trace_r(y)=trace_{rp}(z).$ We evaluate $\alpha$ on $rp$ (which subsumes the evaluation on $r$) and find two nodes $y_i,y_j$ along the path from $y$ to the root, and two nodes $z_i,z_j$ along the path from $z$ to the root, such that:
 $$h_{y_i}=h_{y_j},$$
 $$h_{z_i}=h_{z_j}.$$
Call these two values $h_1$ and $h_2$ respectively. There are several cases to consider, depending on whether and how these paths overlap.  We will just detail the argument in the most complicated case, where $z_i$ is somewhere along the path from $y_i$ to $y_j.$

We thus can write
$$r={\bar r}p_1q_1,$$
where ${\bar r}$ is the tree rooted at $y_i$ and ${\bar r}p_1$ the tree rooted at $y_j.$
We similarly write
$$rp={\bar s}p_2q_2,$$
where ${\bar s}$ is the tree rooted at $z_i$ and ${\bar s}p_2$ is the tree rooted at $z_j.$  Note that because of the overlap, we can write $p_1=cd,$ $p_2=de,$ for contexts $c,d,e,$
and that we have $\alpha(r)=h_1,$ $h_1\alpha(c)=h_2,$ $h_2\alpha(d)=h_1,$ and $h_1\alpha(e)=h_2.$  We insert a new copy of $p_2,$ giving the forest
$$s=rcdedeq_2.$$
Now note that this has the same value $h_2\alpha(q_2)$ under $\alpha$ as $rp,$ and that
$$r_1=rcdedq_1,$$
has the same value $h_1\alpha(q_1)$ under $\alpha$ as $r.$  We can write $s=r_1q,$ where $q$ is a context.  Observe, too, that because $c$ and $e$ were extracted from corresponding segments of equal traces, we have again not introduced any new $(d+1)$-traces in passing from $r$ to $r_1$ that we did not similarly introduce in passing from $rp$ to $s,$ and vice-versa. In the case when the two paths do not overlap, we simply pump $p_1$ and $p_2$ separately, just as we did in part (i). In this manner we remove one by one every $(d+1)$-trace on which the $r$ and $rp$ differ, exactly as in part (i) of the proof.

  \end{document}